\documentclass[10pt,conference]{IEEEtran}
\IEEEoverridecommandlockouts
\usepackage{cite}
\usepackage{amsmath,amssymb,amsfonts}
\usepackage{algorithmic}
\usepackage{graphicx}
\usepackage{textcomp}
\usepackage[table,xcdraw]{xcolor}
\usepackage{multirow}

\usepackage{atbegshi}

\AtBeginShipoutFirst{%
  \put(0,60){\makebox[\textwidth]{\centering This is the accepted version of the paper that will be published in the proceedings of the 22nd IEEE International Conference on}}
  
  \put(0,50){\makebox[\textwidth]{\centering Software Architecture (ICSA 2025). The final published version will be available at IEEE Xplore.}}
  
  \put(0,40){\makebox[\textwidth]{\centering © 2025 IEEE. Personal use of this material is permitted. Permission from IEEE must be obtained for all other uses, in any current or}}
  
  \put(0,30){\makebox[\textwidth]{\centering future media, including reprinting/republishing this material for advertising or promotional purposes, creating new collective works,}}
  
  \put(0,20){\makebox[\textwidth]{\centering or resale or redistribution to servers or lists, or reuse of any copyrighted component of this work in other works.}}

}

\def\BibTeX{{\rm B\kern-.05em{\sc i\kern-.025em b}\kern-.08em
    T\kern-.1667em\lower.7ex\hbox{E}\kern-.125emX}}
    
\begin{document}

\def\code#1{\texttt{#1}}

%\title{\textsc{SecuRe}: A Security Architecture Recommender}
\title{\textsc{SecuRe} - An Approach to Recommending Security Design Patterns}

\author{\IEEEauthorblockN{Alex R. Sabau}
\IEEEauthorblockA{\textit{Research Group Software Construction} \\
\textit{RWTH Aachen University}\\
Aachen, Germany \\
ORCID: 0000-0002-8808-7192}
\and
\IEEEauthorblockN{Dominik Lammers}
\IEEEauthorblockA{\textit{RWTH Aachen University}\\
Aachen, Germany \\
dominik.lammers@rwth-aachen.de}
\and
\IEEEauthorblockN{Horst Lichter}
\IEEEauthorblockA{\textit{Research Group Software Construction} \\
\textit{RWTH Aachen University}\\
Aachen, Germany \\
ORCID: 0000-0002-3440-1238}
}

\maketitle

\begin{abstract}
Security is an important quality of software systems, but there is a huge lack of security experts. To overcome this gap, we aim to make security design knowledge reusable for architects by proposing the \textsc{SecuRe} recommendation approach to secure software design. It lifts design patterns and knowledge engineering concepts to security-related design recommendations for software architectures. This paper presents the central concepts of this approach, the overall recommendation process, and the first results from an initial case study. 
\end{abstract}


\section{Introduction \& Motivation}
Security is an important quality of software systems due to the immense costs induced by security issues \cite{Matulevicius.2017}. Therefore, security experts capable of assessing and ensuring the security of software systems are needed. However, the lack of security experts is not expected to change anytime soon \cite{Furnell.2020}. Moreover, the increased complexity induced by security solutions is considered a major challenge \cite{Deloitte.2021}. To cope with this, we propose \textsc{SecuRe}, a security recommender approach that supports architects by providing design recommendations for security requirements. As a motivation, consider the following story, originally presented by Felfernig and Burke \cite{Felfernig.2008}:\\ 

%Before introducing the \textsc{SecuRe} approach, we would like to motivate you with a short story, originally presented by Felfernig and Burke \cite{Felfernig.2008}:\\

\textit{Ana wants to adapt the architecture of a planned software to satisfy a security requirement. 
She enters the security requirement into her security recommender.
Based on this requirement and further knowledge of the software, such as regulatory policies, the recommender proposes ``Passkey Authentication'' and ``OpenID Connect (OIDC)'' as feasible solutions. After comparing the two options, Ana decides on OIDC. 
Then, she answers a few questions, such as whether a fallback identity provider shall be used.
She receives feedback that solutions that include a fallback identity provider are not viable due to the necessity of a single identity provider.   
She accepts this proposal, and the system recommends the OIDC patterns ``OIDC Client Secret'' and ``OIDC Private Key JWT''. 
Ana asks for more information and learns that the first pattern leads to a better performance but is less secure than the second.
Based on the given system performance requirement, she chooses the ``OIDC Client Secret'' pattern.}\\
\\
\indent This story describes a use case of a constraint-based recommender system (CBRS). These systems use knowledge bases that can derive suitable recommendation items based on their knowledge of the item domain \cite{Jannach.2010}. Our research question sums up the central aspects of this motivating example:

\noindent \textbf{RQ:} How can a CBRS-based recommendation approach be designed to support architects in designing secure software systems?

%\noindent \textbf{RQ:} How can a CBRS be designed that supports architects in designing secure software systems?

\section{Related Work}
Recommender systems have been the subject of research for many years. Recommendation approaches exist for object-oriented design patterns \cite{Gomes.2002 , Kung.2003}. Pescador and de Lara \cite{Pescador.2016} propose a DSL recommender using meta-model design patterns, a concept ontology, and natural language analysis for meta-model generation. Brandner and Weinreich \cite{Brandner.2019} use historical design decisions to recommend future design choices. Kögel \cite{Kögel.2017} developed a recommender for model-driven software development, suggesting model editions based on past changes. The approach of Kuschke et al. \cite{Kuschke.2013} recommends UML model completion at the time of modeling. Sen et al. \cite{Sen.2010} propose a methodology for recommending completions of incomplete models in domain-specific modeling languages. Then, much research has been done into security patterns, but their adoption in practice remains limited \cite{van_den_Berghe.2022}. To the best of our knowledge, a CBRS that generates design recommendations for security-related properties does not yet exist. Further, \textsc{SecuRe} takes a novel approach to reuse knowledge about security patterns and uses a proven concept from recommender theory to calculate recommendations for these patterns.

\section{Concepts of Security Design Knowledge}

\begin{figure}[t]
    \centering
    \includegraphics[width=0.7\linewidth]{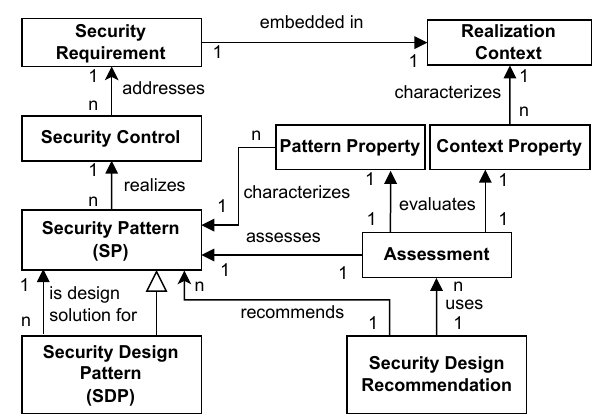}
    \caption{Security Design Recommendation Concept Model}
    \label{fig:concepts}
\end{figure}

In the following, we present the central concepts the \textsc{SecuRe} approach is based on, depicted in Figure \ref{fig:concepts}. A \textit{security requirement} specifies a security condition to be met by a system, following the security-by-design paradigm. Well-known security controls exist to satisfy security requirements. Based on NIST \cite{FIPS.200_2006}, a \textit{security control} is an action, procedure, technique, or other measure that reduces the vulnerability of a software system. Examples are authentication, authorization, and DDoS prevention. Since various options exist for each security control to be realized, we use the concept of a security pattern, introduced by Yoder and Barcalow \cite{Yoder.1998},   to group these options. A \textit{security pattern} (SP) is a reusable solution for a security control on the conceptual level without concrete implementation details. Each SP can be characterized by a set of specific \textit{pattern properties}. A \textit{security design pattern} (SDP) concretizes an SP by defining a reusable design solution for it, thus realizing a security control on the design level. An SDP defines the detailed structure and behavior of components that implement a security control according to the conceptual solution. SPs and SDPs offer a \textit{description} explaining the application and the pattern's mode of action. In conclusion, security controls, SPs, and SDPs form a hierarchy. For each security control, there are multiple SPs on the conceptual level. For each SP, there are multiple SDPs on the design level.

These concepts can be explained clearly using the example of the security control authentication (AuthN). There are several ways to realize it, e.g., \textit{password-based} or \textit{passkey-based AuthN}. Each option corresponds to an SP; they represent conceptual solutions of AuthN. An example of a pattern property of these SPs is the \textit{authentication strength}, as password-based AuthN is considered weaker than passkey-based AuthN. Further, there are several ways to implement the password-based AuthN SP. For instance, a centralized password-based AuthN design would be feasible in a monolithic architecture. In contrast, it should follow a decentralized design in a decentralized architecture such as blockchain \cite{Szalachowski.2021}. \textit{Centralized} and \textit{decentralized password-based AuthN} are both SDPs of the password-based AuthN SP. An example of a pattern property for these SDPs is the \textit{password reset mechanism} used.

Further, a security requirement is always embedded in a \textit{realization context}, which defines additional conditions to be met. Regulatory policies, for example, define conditions that a security control must comply with. We refer to this kind of condition as a \textit{context property}. Pattern and context properties need to be evaluated, yielding an \textit{assessment} of an SP or an SDP to realize a security control for a given security requirement and its realization context. A \textit{security design recommendation} is based on these assessments.

As SDPs are just special SPs, we will talk about SPs only for the sake of simplicity in the rest of the paper.

\section{Knowledge Elements of \textsc{SecuRe}}
%Based on the concepts presented, we briefly introduce the central knowledge elements of \textsc{SecuRe} using the usual CBRS terminology in this section. A detailed introduction to the CBRS theory can be found in Jannach et al. \cite{Jannach.2010}. A KB for \textsc{SecuRe} consists of the following variable and constraint sets:
Based on the concepts presented, we introduce the central knowledge elements of \textsc{SecuRe} using the usual CBRS terminology in this section. A detailed introduction to the CBRS theory can be found in Jannach et al. \cite{Jannach.2010}.

\textsc{SecuRe} uses a set of knowledge bases (KB): one KB for each security control and one KB for each SP. Each KB consists of the following variable and constraint sets:

\begin{itemize}
  \item \textit{Context properties:} Contains variables that represent the context properties of a realization context
  \item \textit{Pattern properties:} Contains variables that represent the pattern properties of an SP
  \item \textit{Contextual constraints:} Contains rules for valid combinations of context properties
  \item \textit{Filter conditions:} Contains rules that represent relationships between context properties and pattern properties
  \item \textit{Valid SPs:} Contains rules for valid SPs since not all combinations of pattern properties are valid SPs
\end{itemize}

\textsc{SecuRe} solves constraint satisfaction problems (CSP) using constraint solvers. A constraint solver has to find valid SPs that meet all contextual constraints and filter conditions. Solving the CSP results in the list of all potentially feasible SPs \cite{Jannach.2010}. Then, \textsc{SecuRe} applies the multi-attribute utility theory (MAUT) to calculate a recommendation score on these SPs. 

In essence, MAUT ranks alternatives by assigning weights and utilities to certain criteria \cite{Jannach.2010}. These criteria are---often conflicting---system properties that the architect must optimize in their decision-making, such as ``performance'', ``security'', or ``costs''. \textsc{SecuRe} encodes weights on these properties in relation to certain context properties. It uses specific weights derived from experience and knowledge engineering, similar to COCOMO \cite{Boehm.2009}. These weights affect how suitable \textsc{SecuRe} considers an SP to be in a given realization context. For instance, for the system property ``performance'', weights can be defined for the context property ``required security level''. If its weight is low, ``performance'' will be ranked higher; if it is high, ``performance'' will be ranked lower. This rank affects the suitability of SPs and, thus, their recommendation scores: If ``performance'' is ranked high, SPs with better performance are more feasible, i.e., their good performance has a stronger positive effect on their recommendation scores. However, these SPs are less attractive if ``performance'' is ranked lower, as other properties can be optimized instead, resulting in a less positive or even negative effect on the recommendation scores.

\section{Recommendation Process with \textsc{SecuRe}}

\begin{figure}[t]
    \centering
    \includegraphics[width=\linewidth]{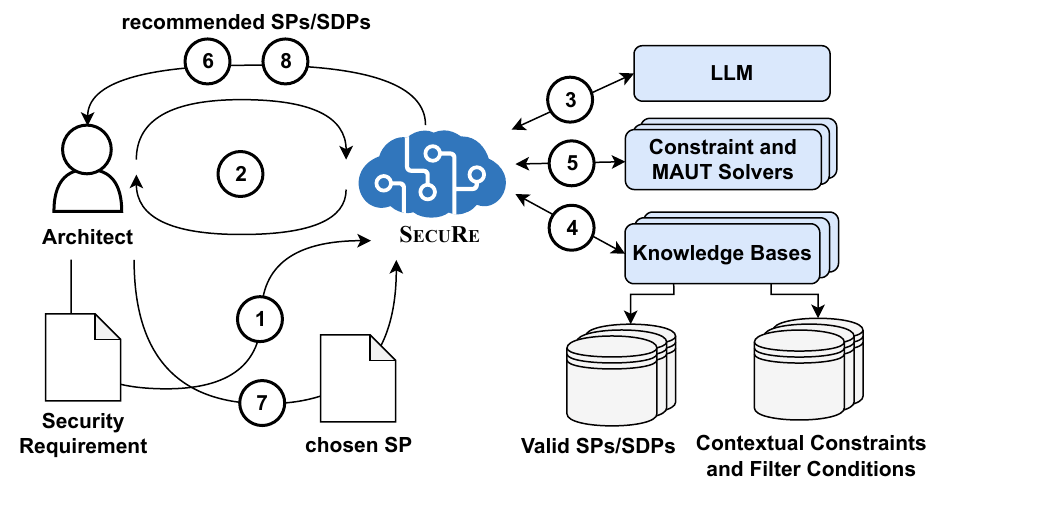}
    \caption{The \textsc{SecuRe} recommendation process}
    \label{fig:approach_high_level}
\end{figure}

\textsc{SecuRe}'s recommendation process is depicted in Figure \ref{fig:approach_high_level}. It contains the following eight steps: 

\textbf{1:} The architect provides a security requirement.

\textbf{2:} \textsc{SecuRe} uses a question and answer (Q\&A) loop to define the realization context, i.e., all context property values.

\textbf{3:} To improve the precision of the realization context, the architect can also ask questions in the Q\&A loop. \textsc{SecuRe} uses a Large Language Model (LLM) to answer them.

\textbf{4:} After the realization context is set, \textsc{SecuRe} accesses its KBs. Each KB stores the valid SPs or SDPs, their contextual constraints, and filter conditions. 

\textbf{5:} Based on the context properties, valid SPs, contextual constraints, and filter conditions, \textsc{SecuRe} solves the CSP using a constraint solver. Then, it applies MAUT to calculate a recommendation score of each SP. This results in a list of all potentially feasible SPs and their recommendation scores.

\textbf{6:} \textsc{SecuRe} returns the description of these SPs, ordered by the recommendation scores, along with reasoning for each recommendation to ensure transparency and clarity.

\textbf{7:} After analyzing the SP descriptions, the architect decides on an SP and provides it as input to \textsc{SecuRe}. Now, \textsc{SecuRe} repeats steps 2 to 5 to find suitable SDPs to the selected SP.

\textbf{8:} This time, \textsc{SecuRe} returns a list of potentially feasible SDPs and provides the SDP descriptions to the architect.

\section{Initial Case Study}

We evaluated \textsc{SecuRe} in an initial case study, limited to SP recommendations for the security control AuthN only, to get first insights into its feasibility and performance. To this end, we modeled a total of six AuthN SPs characterized by five pattern properties, and eight realization contexts characterized by six context properties. Further, we defined three filter conditions on three context properties and three pattern properties. The pattern properties were the following:

\begin{itemize}
    \item \code{AuthN-strength}: the AuthN strength of the SP
    \item \code{AuthN-usablty}: the AuthN usability of the SP
    \item \code{costs}: the overall costs incurred by the SP
    \item \code{dev-bind}: the SP requires device(s) bound to the user
    \item \code{add-dev}: the SP needs additional non-personal device
\end{itemize}

\noindent With these, we have defined the following six AuthN patterns; their exact characterizations are presented in Table \ref{tab:SPs}:

\begin{itemize}
    \item \code{password}: common password-based AuthN
    \item \code{key-stretch}: password AuthN with key-stretching
    \item \code{hrdw-token}: hardware-token-based AuthN
    \item \code{passkey}: passkey-based AuthN 
    \item \code{biom-device}: biometric AuthN on user devices
    \item \code{biom-profile}: biometric AuthN with user profiles  
\end{itemize}

\noindent The context properties were defined as follows:

\begin{itemize}
    \item \code{sec-lev}: the required level of security
    \item \code{use-lev}: the required level of usability
    \item \code{budget}: the amount of budget that can be invested
    \item \code{no-users}: the number of AuthN users
    \item \code{intern-extern}: AuthN users are internal or external
    \item \code{shared-device}: AuthN users must share devices
\end{itemize}

To calculate the recommendation scores, we applied MAUT and selected ``usability'' and ``costs'' as the properties to be optimized. We defined weights on them for the context properties \code{sec-lev}, \code{use-lev}, \code{budget}, and \code{no-users}. E.g., low \code{budget} led to a low ``usability'' and high ``costs'' weight; in low-budget use cases, cheap SPs are more favorable than usable ones.

With the context properties, we defined eight realization contexts (RC). Based on their characterizations, our experience, and insights from the literature, we determined the expected SP recommendations for each RC. All RCs and the expectations determined for them are shown in Table \ref{tab:evaluation_config}. In the following, the expectations are briefly discussed:

\begin{itemize}
    \item In RCs with a high required level of security, \code{password} should not be recommended (RC3, RC6).
    \item In RCs with external users, \code{hrdw-token} should not be recommended (RC7, RC8). 
    \item In RCs with shared devices, \code{passkey}, and \code{biom-device} should not be recommended (RC6, RC8), \code{biom-profile} should still be recommended.
    \item In RCs with a high budget and a low number of users, the pattern property \code{AuthN-usablty} should be weighted higher, and the pattern property \code{costs} lower, i.e., more usable SPs should be recommended highest: \code{passkey}, \code{biom-device}, and \code{biom-profile} (RC1, RC2, RC3, RC7).
    \item In RCs with low budget and high number of users, the opposite is expected. Cheaper SPs should be recommended highest: \code{password} being the cheapest SP (RC4, RC5).
    \item In RCs with low budget and low number of users, i.e., overall lower costs, more usable SPs might be recommended highest: \code{password} or \code{biom-profile} (RC8).
    \item The \code{hrdw-token} and \code{key-stretch} SPs should never be recommended highest. The first has limited usability and high costs, the latter is just slightly securer but also more expensive than \code{password}.
\end{itemize}

\begin{figure*}[t]
    \includegraphics[width=\textwidth]{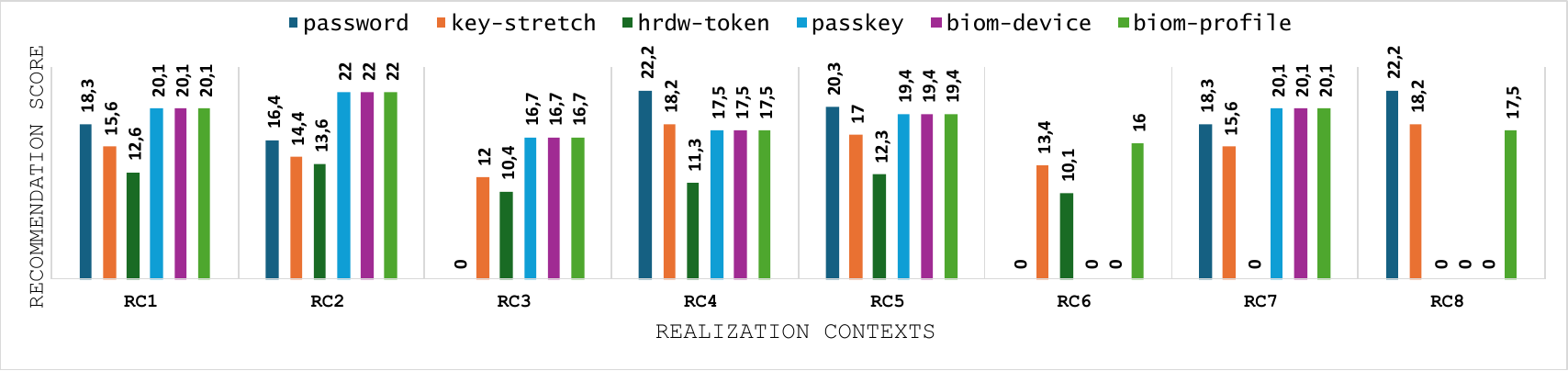}
    \caption{Recommendation scores for the defined realization contexts.}
    \label{fig:evaluation}
\end{figure*}

The calculated recommendation scores of the six AuthN SPs for all eight RCs are depicted in Figure \ref{fig:evaluation}. Overall, the evaluation results obtained in this case study are promising, as the recommended SPs align with our expectations.

\begin{table}[t]
\caption{Properties of the AuthN SPs used in the case study}
\label{tab:SPs}
\centering
    \resizebox{\linewidth}{!}{
\begin{tabular}{l|l|l|l|l|l}
                        & \code{AuthN} & \code{AuthN} & \code{costs}  & \code{dev}  & \code{add}     \\ 
                        & \code{strength} & \code{usablty} &  & \code{bind} & \code{dev}                     \\ \hline
\code{password} & low & low & low & agnostic & no            \\ \hline
\code{key-stretch} & medium & low & medium & agnostic& no    \\ \hline
\code{hrdw-token}   & medium & medium & high & agnostic & yes \\ \hline
\code{passkey}  & high & high & high & bound & no           \\ \hline
\code{biom-device}   & medium & high & high & bound  & no               \\ \hline
\code{biom-profile}  & medium & high & high & agnostic   & no          
\end{tabular}
    }
\end{table}

%%% IF FIGURE IS PRINTED ON PAGE 5, PUT TABLE BELOW TO HERE

\section{Conclusion and Future Work}
With \textsc{SecuRe}, we presented a security recommender approach to support architects' decision-making for secure software systems. \textsc{SecuRe} is, to our knowledge, the first of its kind. We showed its feasibility in an initial case study with promising results. \textsc{SecuRe} not only supports architects in designing security solutions but also lays the foundation for reusing security design knowledge, marking our results of great value for practitioners and researchers.

We plan to extend \textsc{SecuRe} to include SDP recommendations for future work. Further, we will apply more knowledge engineering to encode and cover a broader range of context and security pattern properties. We are also currently working on a methodological approach to systematically extract knowledge for context and pattern properties using LLMs. Lastly, we will conduct further case studies and an expert study to evaluate \textsc{SecuRe} in practice.

\begin{table}[t]
\caption{Realization contexts and expected recommendations}
\label{tab:evaluation_config}
\centering
    \resizebox{\linewidth}{!}{
\begin{tabular}{l|llllll}
\multicolumn{1}{l|}{\begin{tabular}[c]{@{}l@{}}\textbf{RCs}\\ \textbf{Exp. SPs}\end{tabular}} & \multicolumn{1}{l|}{\begin{tabular}[c]{@{}l@{}}\code{sec-lev}\end{tabular}} & \multicolumn{1}{l|}{\begin{tabular}[c]{@{}l@{}}\code{use-lev}\end{tabular}} & \multicolumn{1}{l|}{\code{budget}} & \multicolumn{1}{l|}{\begin{tabular}[c]{@{}l@{}}\code{no}\\ \code{users}\end{tabular}} & \multicolumn{1}{l|}{\begin{tabular}[c]{@{}l@{}}\code{intern}\\ \code{extern}\end{tabular}} & \begin{tabular}[c]{@{}l@{}}\code{shared}\\ \code{device}\end{tabular} \\ \hline \hline
 & \multicolumn{1}{l|}{low} & \multicolumn{1}{l|}{low} & \multicolumn{1}{l|}{high} & \multicolumn{1}{l|}{low} & \multicolumn{1}{l|}{internal} & no \\ \cline{2-7}
\multirow{-2}{*}{\textbf{\begin{tabular}[c]{@{}l@{}}RC1\\  best SPs:\end{tabular}}} & \multicolumn{6}{l}{\code{passkey}, \code{biom-device}, \code{biom-profile}} \\ \hline
\rowcolor[HTML]{EFEFEF} 
\cellcolor[HTML]{EFEFEF} & \multicolumn{1}{l|}{\cellcolor[HTML]{EFEFEF}low} & \multicolumn{1}{l|}{\cellcolor[HTML]{EFEFEF}high} & \multicolumn{1}{l|}{\cellcolor[HTML]{EFEFEF}high} & \multicolumn{1}{l|}{\cellcolor[HTML]{EFEFEF}low} & \multicolumn{1}{l|}{\cellcolor[HTML]{EFEFEF}internal} & no \\ \cline{2-7}
\rowcolor[HTML]{EFEFEF} 
\multirow{-2}{*}{\cellcolor[HTML]{EFEFEF}\textbf{\begin{tabular}[c]{@{}l@{}}RC2\\  best SPs:\end{tabular}}} & \multicolumn{6}{l}{\cellcolor[HTML]{EFEFEF}\code{passkey}, \code{biom-device}, \code{biom-profile}} \\ \hline
 & \multicolumn{1}{l|}{high} & \multicolumn{1}{l|}{low} & \multicolumn{1}{l|}{high} & \multicolumn{1}{l|}{low} & \multicolumn{1}{l|}{internal} & no \\ \cline{2-7}
\multirow{-2}{*}{\textbf{\begin{tabular}[c]{@{}l@{}}RC3\\  best SPs:\end{tabular}}} & \multicolumn{6}{l}{\code{passkey}, \code{biom-device}, \code{biom-profile}} \\ \hline
\rowcolor[HTML]{EFEFEF} 
\cellcolor[HTML]{EFEFEF} & \multicolumn{1}{l|}{\cellcolor[HTML]{EFEFEF}low} & \multicolumn{1}{l|}{\cellcolor[HTML]{EFEFEF}low} & \multicolumn{1}{l|}{\cellcolor[HTML]{EFEFEF}low} & \multicolumn{1}{l|}{\cellcolor[HTML]{EFEFEF}high} & \multicolumn{1}{l|}{\cellcolor[HTML]{EFEFEF}internal} & no \\ \cline{2-7}
\rowcolor[HTML]{EFEFEF} 
\multirow{-2}{*}{\cellcolor[HTML]{EFEFEF}\textbf{\begin{tabular}[c]{@{}l@{}}RC4\\  best SPs:\end{tabular}}} & \multicolumn{6}{l}{\cellcolor[HTML]{EFEFEF}\code{password}} \\ \hline
 & \multicolumn{1}{l|}{low} & \multicolumn{1}{l|}{high} & \multicolumn{1}{l|}{low} & \multicolumn{1}{l|}{high} & \multicolumn{1}{l|}{internal} & no \\ \cline{2-7}
\multirow{-2}{*}{\textbf{\begin{tabular}[c]{@{}l@{}}RC5\\  best SPs:\end{tabular}}} & \multicolumn{6}{l}{\code{password}} \\ \hline
\rowcolor[HTML]{EFEFEF} 
\cellcolor[HTML]{EFEFEF} & \multicolumn{1}{l|}{\cellcolor[HTML]{EFEFEF}high} & \multicolumn{1}{l|}{\cellcolor[HTML]{EFEFEF}high} & \multicolumn{1}{l|}{\cellcolor[HTML]{EFEFEF}low} & \multicolumn{1}{l|}{\cellcolor[HTML]{EFEFEF}high} & \multicolumn{1}{l|}{\cellcolor[HTML]{EFEFEF}internal} & yes \\ \cline{2-7}
\rowcolor[HTML]{EFEFEF} 
\multirow{-2}{*}{\cellcolor[HTML]{EFEFEF}\textbf{\begin{tabular}[c]{@{}l@{}}RC6\\  best SPs:\end{tabular}}} & \multicolumn{6}{l}{\cellcolor[HTML]{EFEFEF}\code{biom-profile}} \\ \hline
 & \multicolumn{1}{l|}{low} & \multicolumn{1}{l|}{low} & \multicolumn{1}{l|}{high} & \multicolumn{1}{l|}{low} & \multicolumn{1}{l|}{external} & no \\ \cline{2-7}
\multirow{-2}{*}{\textbf{\begin{tabular}[c]{@{}l@{}}RC7\\  best SPs:\end{tabular}}} & \multicolumn{6}{l}{\code{passkey}, \code{biom-device}, \code{biom-profile}} \\ \hline
\rowcolor[HTML]{EFEFEF} 
\cellcolor[HTML]{EFEFEF} & \multicolumn{1}{l|}{\cellcolor[HTML]{EFEFEF}low} & \multicolumn{1}{l|}{\cellcolor[HTML]{EFEFEF}low} & \multicolumn{1}{l|}{\cellcolor[HTML]{EFEFEF}low} & \multicolumn{1}{l|}{\cellcolor[HTML]{EFEFEF}low} & \multicolumn{1}{l|}{\cellcolor[HTML]{EFEFEF}external} & yes \\ \cline{2-7}
\rowcolor[HTML]{EFEFEF} 
\multirow{-2}{*}{\cellcolor[HTML]{EFEFEF}\textbf{\begin{tabular}[c]{@{}l@{}}RC8\\  best SPs:\end{tabular}}} & \multicolumn{6}{l}{\cellcolor[HTML]{EFEFEF}\code{password} or \code{biom-profile}}
\end{tabular}
}
\end{table}

\bibliographystyle{IEEEtran}
\bibliography{IEEEabrv,Bibliography}

\end{document}